\documentclass[prb,aps]{revtex4}
\usepackage{epsfig}

\begin{document}
\draft 
\title{Electron injection in a nanotube: noise correlations and
entanglement}   
\author{A. Cr\'epieux$^{1,2}$, R.
Guyon$^{1,2}$, P. Devillard$^{1,3}$ and T. Martin$^{1,2}$}     
\address{ $^1$ Centre de Physique  Th\'eorique,  
 Case 907 Luminy, 13288 Marseille Cedex 9, France} 
\address{$^2$ Universit\'e de la M\'editerran\'ee,  
13288 Marseille Cedex 9, France}
\address{$^3$ Universit\'e de Provence, 13331 Marseille Cedex 
03, France}

\begin{abstract} 
Transport through a metallic carbon nanotube is considered,
where electrons are injected in the bulk by a scanning tunneling microscope tip. 
The charge current and noise are computed both in
the  absence and in the presence of one dimensional Fermi liquid 
leads. For an infinite homogeneous nanotube, the shot noise
exhibits effective charges different from the electron 
charge. Noise correlations between both ends of the 
nanotube are positive, and occur to second order
only in the tunneling amplitude.
The positive correlations are symptomatic of an entanglement 
phenomenon between quasiparticles moving right and left 
from the tip. This entanglement involves many body states
of the boson operators which describe the 
collective excitations of the Luttinger liquid.  
\end{abstract}

\maketitle

\section{Introduction}

Over the years, the study of current noise and noise correlations
has become a respected and useful diagnosis for transport
measurements on mesoscopic conductors. Theoretically, noise was first computed mostly for
non--interacting systems 
\cite{blanter_buttiker}. However, it soon became clear  that
low frequency noise could be used to isolate the  quasiparticle
charge \cite{kane_fisher_wen_fendley,glattli_reznikov}
and to study the statistical correlations 
\cite{saleur_weiss,safi_devillard_martin}
in specific quasi one--dimensional correlated electron
systems, such as the edge waves in the quantum Hall effect.
In these
chiral Luttinger liquids, the
charge of the collective excitations along the edges corresponds
to the electron charge multiplied by the filling factor. 

Attention is
now turning towards conductors -- individual nano-objects --
which occur naturally, and which can be connected to current/voltage 
probes in order to perform a transport experiment. 
The crucial advantage of such nano--objects is that
they are essentially free of defects and in some
circumstances they have an inherent one dimensional
character. Carbon nanotubes constitute the archetype of 
such 1D nano-objects: single wall armchair nanotubes have 
metallic behavior,
with two propagating modes at the Fermi level. Incidentally, 
electronic correlations are known to play an important 
role in such systems. Carbon nanotubes seem to constitute good
candidates to study Luttinger liquid behavior. In
particular, their tunneling density of states -- and thus the
tunneling $I(V)$ characteristics is known to have a power law
behavior \cite{kane_balents_fisher,mceuen,egger_review} in
accordance  with Luttinger liquid theory.

\begin{figure}  
\epsfxsize 8 cm  
\centerline{\epsffile{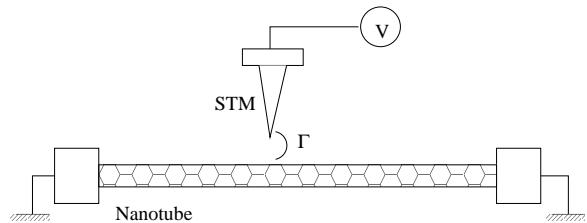}}  
\medskip  
\caption{\label{fig1} Schematic configuration of the
nanotube--STM device: electrons are injected from the tip
at $x=0$: current is measured at both nanotube ends, which
are set to the ground.}    
\end{figure} 

Luttinger models for nanotubes differ significantly from their
quantum Hall effect counterpart, because of their non-chiral
character.  Forward and backward fields describing collective
excitations effectively mix, because the interactions between
electrons are spread along the whole length of the nanotube. For
this reason, a straightforward transposition of the  results
obtained for chiral edge system proves difficult. Nevertheless,
non--chiral Luttinger liquids can be described with chiral 
fields \cite{ines_annales,pham}. Such chiral fields correspond 
to excitations with anomalous (non-integer) charge, which has
eluded  detection so far. 

In the present work, we propose an experimental geometry which
allows to probe directly the underlying charges of the
collective excitations. The setup consists of a nanotube
whose bulk is contacted by a scanning tunneling microscope (STM) tip which injects electrons,
while both extremities of the nanotube collect the current
(Fig. \ref{fig1}). The current, the noise and the noise
correlations are computed, and the effective charges are
determined by comparison with the Schottky formula 
\cite{shottky} for an
``infinite'' nanotube, the striking result is that noise
correlations contribute to second order in the electron
tunneling, in sharp contrast with a fermionic 
system which requires fourth order. The noise correlations are
then positive, because the tunneling electron wave function is
split in two counter propagating modes of the collective
excitations in the  nanotube. We conjecture that in the presence of 1D Fermi
 liquid leads, modeled as in Ref. \cite{safi_maslov}, the absence of renormalization/interaction
effects of the nanotube is recovered.

A recent two terminal experiment studied the
current--current fluctuations in ropes of nanotubes \cite{roche}.
There, it is  pointed out that the strong reduction of the 
low frequency noise cannot be  understood within the context of
scattering  theory \cite{martin_landauer_buttiker}.
Naive comparison with existing non-chiral Luttinger liquid models
\cite{pham} would imply an interaction parameter much inferior 
to the free electron case. Also, we mention that other 
multi-terminal geometries where a nanotube or a 
one--dimensional wire is attached to more than two leads, have
been considered \cite{blanter,imura,trauzettel,chen,pham2}. 
Our proposal deals with the same geometry as Ref \cite{imura}, 
where a renormalization analysis identified the exponents
of the current voltage characteristics. However, here the 
emphasis is put on the low frequency
current fluctuation spectrum, both for the 
autocorrelation and the cross correlations between the two ends
of the nanotube.   

The paper is organized as follows: the Hamiltonian of 
our setup is specified in the next section, followed by a
general non-equilibrium scheme based on the Keldysh formalism to
study transport in this device, which is independent of the type
of leads chosen (Sect. 3). Results for a nanotube connected to
leads are then presented in Sect. 4. A connection 
with the  effective charges of Refs.
\cite{ines_annales,pham,imura} is established
in Sect. 5.
  
\section{Model Hamiltonian}

The transport geometry (Fig.
\ref{fig1}) implies
tunneling from the tip (normal or ferromagnetic metal)
to the nanotube, and subsequent propagation of collective 
excitations along 
the nanotube. 
In the absence of tunneling, 
the Hamiltonian is thus simply the sum
of the nanotube Hamiltonian, described by a two mode
Luttinger liquid, together with the tip Hamiltonian. Using the
standard conventions \cite{egger_gogolin}, the operator
describing  an electron with spin $\sigma$ moving along the
direction $r$, from mode $\alpha$ is specified  in terms of a
bosonic field:   
\begin{equation} 
\Psi_{r\alpha \sigma}(x,t)={1\over \sqrt{2\pi a}}
e^{i \alpha k_Fx + i r q_Fx + i \varphi_{r\alpha \sigma}(x,t)}
~,\label{fermion_nanotube}
\end{equation}
with $a$ a short distance cutoff, $k_F$ the Fermi momentum,
$q_F$ the momentum mismatch associated with the two modes, and
the convention $r=\pm$, $\alpha=\pm$ and $\sigma=\pm$ are chosen
for the direction of propagation, for the nanotube branch, and
for the spin orientation.   
It is convenient to express this
bosonic phase  in terms of the  conventional non-chiral
Luttinger liquid fields  $\theta_{j\delta}$ and
$\phi_{j\delta}$, with $j\delta\in\{c+,c-,s+,s-\}$ identifying
the charge/spin  and total/relative fields:  
\begin{eqnarray}
\varphi_{r\alpha\sigma}(x,t)=\sqrt{\frac{\pi}{2}}
\sum_{j\delta}h_{\alpha\sigma
j\delta}(\phi_{j\delta}(x,t)+r\theta_{j\delta}(x,t))
~,\end{eqnarray} 
with $h_{\alpha\sigma c+}=1$, $h_{\alpha\sigma
c-}=\alpha$,  $h_{\alpha\sigma s+}=\sigma$ et $h_{\alpha\sigma
s-}=\alpha\sigma$. $\theta_{j\delta}$ and $\phi_{j\delta}$
are dual non-chiral fields. 
A plausible alternative would have been
to express $\varphi_{r\alpha \sigma}$ in terms of the 
chiral Luttinger liquid fields. However, the present 
choice will be simpler later on when dealing with inhomogeneous
Luttinger liquids (in order to include the leads),
as the Green's functions for $\theta_{j\delta}$,
$\phi_{j\delta}$ are known. The Hamiltonian which describes 
the collective excitations in the nanotube has the standard form:
\begin{eqnarray}
H=\frac{1}{2}\sum_{j\delta}\int_{-\infty}^{\infty}dx\bigg(v_{j\delta}K_{j\delta}
(\partial_x\phi_{j\delta}(x,t))^2
+\frac{v_{j\delta}}{K_{j\delta}}(\partial_x\theta_{j\delta}(x,t))^2\bigg)
~,\end{eqnarray}
with an interaction
parameter $K_{j\delta}$ and velocity $v_{j\delta}$.

For the STM tip, one assumes for simplicity that only one
electronic mode  couples to the nanotube.  
The tip can thus be described by a semi-infinite 
Luttinger liquid, as in Kondo type problems. This turns out to
be convenient in this problem where both bosonized nanotube 
fermions operators and tip fermions operators intervene. For
the sake of generality, we allow the two spin components of
the tip fields to have different Fermi velocities $u_F^\sigma$,
which allows to treat the case  of a ferromagnetic metal. The
fermion operator at the tip location $x=0$ is then:
\begin{equation}    c_{\sigma}(t)=
\frac{1}{\sqrt{2\pi a}}
e^{i\tilde{\varphi}_{\sigma}(t)}~.
\label{fermion_stm}
\end{equation}
Here, $\tilde{\varphi}_{\sigma}$ is the chiral Luttinger liquid
field, whose Keldysh Green's function at
$x=0$ is given by \cite{chamon_freed}: \begin{eqnarray}
g_{\sigma(\eta_1\eta_2)}(t_1,t_2)&\equiv&
\langle T_K\{\tilde{\varphi}_{\sigma}(t_1^{\eta_1})
\tilde{\varphi}_{\sigma}(t_2^{\eta_2})\}\rangle\nonumber\\
&=&-\frac{1}{2\pi}
\ln\left\{1+i[(\eta_1+\eta_2){\mathrm sgn}(t_1-t_2)-(\eta_1-\eta_2)] 
u_F^{\sigma}(t_1-t_2)/2a\right\}~, \end{eqnarray} 
where $\eta_{1,2}=\pm$ refer to the upper or lower 
branch of the Keldysh contour. 

The tunneling Hamiltonian is a standard hopping term:
\begin{eqnarray}
H_T(t)=\sum_{\varepsilon r \alpha \sigma}
\Gamma_{r\alpha\sigma}^{(\varepsilon)}(t)
[\Psi_{r\alpha\sigma}^\dagger(0,t)c_{\sigma}(t)]^{(\varepsilon)}
~.\label{tunnel_hamiltonian}
\end{eqnarray}
Here the superscript $(\varepsilon)$ leaves either the
operators in bracket unchanged ($\varepsilon=+$), or transforms
them into their Hermitian conjugate ($\varepsilon=-$). The 
voltage bias between the tip and the nanotube is included using
the Peierls substitution: the hopping amplitude
$\Gamma_{r\alpha\sigma}^{(\varepsilon)}$ acquires a time
dependent  phase $\exp(i\varepsilon \omega_0 t)$, with the bias
voltage  identified as $V=\hbar\omega_0/e$. We
will use the convention $\hbar\rightarrow 1$. Similarly, 
the tunneling current is defined as: \begin{eqnarray}
I_T(t)=ie\sum_{\varepsilon r\alpha\sigma}\varepsilon 
\Gamma_{r\alpha\sigma}^{(\varepsilon)}(t)
[\Psi_{r\alpha\sigma}^\dagger(0,t)c_{\sigma}(t)]^{(\varepsilon)}
~.\end{eqnarray}

In Eqs. (\ref{fermion_nanotube}) and (\ref{fermion_stm}), we
have omitted the Klein factors which guarantee
the anti-commutation of the 3 types 
of fermions operators -- written in terms of bosonic fields --
for this problem: the two nanotube branches and the STM single
mode. It has been established \cite{safi_devillard_martin,guyon} that  
Klein factors are in principle necessary to treat 
multi-Luttinger system, as illustrated in the computation 
of noise correlations between three edge states in the 
FQHE. In the present work, 
Klein factors can be dropped because we intend to work 
with lowest order perturbation theory. To order $\Gamma^2$, 
statistical correlations between the three Luttinger systems 
do not occur. However, they should show up when calculating 
higher order corrections  ($\Gamma^4$).   

For this problem which implies propagation along the nanotube,  it
is also necessary to compute the  (total) charge and (total) spin
currents using the bosonized fields of Eq. (\ref{fermion_nanotube}):
\begin{eqnarray}
I_\rho(x,t)&=&
ev_F\sum_{r\alpha\sigma}r\Psi^\dagger_{r\alpha\sigma}(x,t)\Psi_{r\alpha\sigma}(x,t)
\nonumber\\
&=&2ev_F\sqrt{\frac{2}{\pi}}\partial_x\phi_{c+}(x,t)
~.
\label{charge_current_operator} \end{eqnarray}
Similarly, we consider the spin current in the $\hat{z}$
direction: 
\begin{eqnarray}
I_{\sigma_z}(x,t)&=&ev_F\sum_{r\alpha\sigma}r\Psi^\dagger_{r\alpha\sigma}(x,t)
\sigma_z \Psi_{r\alpha\sigma}(x,t)\nonumber\\
&=&2ev_F\sqrt{\frac{2}{\pi}}\partial_x\phi_{s+}(x,t)
~.\label{spin_current_operator}
\end{eqnarray}
Note that the contribution from terms containing  $2k_F$
oscillations has been dropped. This is equivalent to requiring
that the current measurement along the nanotube is effectively
a spatial average over a length scale larger than $\lambda_F$.
In practice, $2k_F$ terms are necessary in order to 
establish a connection between current fluctuations
and density fluctuations.  

\section{Non equilibrium transport formalism}

In this section, the general approach used to calculate the
tunneling current and noise, as well as the current and noise
in the nanotube is described. All quantities are computed at
zero temperature for simplicity. The calculation of the tunneling
current and noise is quite similar to  the perturbative results
in  Ref. \cite{kane_fisher_wen_fendley} for the FQHE. Here it is
summarized in order to compare with the nanotube transport
quantities.

\subsection{Tunneling current and noise}

The Keldysh technique is used to compute the average
tunneling current and noise. We adopt the convention that  the
coefficients $\eta,\eta_{1,2}=\pm$ identify the upper/lower
branch of the Keldysh contour:
\begin{eqnarray} \langle
I_T(t)\rangle&=&\frac{1}{2}\sum_\eta\langle T_K\{I_T(t^\eta)
e^{-i\int_K dt_1H_T(t_1)}\}\rangle~, \label{definition_current}\\
S_T(t,t')&=&\frac{1}{2}\sum_\eta\langle
T_K\{I_T(t^\eta)I_T(t'^{-\eta})e^{-i\int_K
dt_1H_T(t_1)}\}\rangle ~,
\end{eqnarray} 
which applies in typical tunneling situations 
where the product of the current averages is of order
$\Gamma^4$.  In order to collect the
lowest order contribution in the tunneling amplitude, the
exponential is expanded to first order for the current, and to
zeroth order for the noise:   \begin{eqnarray} 
\langle I_T(t)\rangle&=&
\frac{e\Gamma^2}{2}\sum_{r\alpha\sigma\varepsilon\eta\eta_1}\eta\varepsilon
\int_{-\infty}^{+\infty}dt_1e^{-i\varepsilon\omega_0(t-t_1)}
\langle
T_K\{\Psi_{r\alpha\sigma}(0,t^{\eta})^{(\varepsilon)}\Psi_{r\alpha\sigma}(0,t_1^{\eta_1})^{(-\varepsilon)}\}\rangle 
\langle
T_K\{c_{\sigma}(t^{\eta})^{(-\varepsilon)}c_{\sigma_1}(t_1^{\eta_1})^{(\varepsilon)}\}\rangle~,
\label{intermediate_tunneling_current}\\
S_T(t,t')&=&
\frac{e^2\Gamma^2}{2}\sum_{r\alpha\sigma\varepsilon\eta}
e^{-i\varepsilon\omega_0(t-t')}
\langle
T_K\{\Psi_{r\alpha\sigma}^+(0,t^{\eta})^{(\varepsilon)}
\Psi_{r\alpha\sigma}^+(0,t'^{-\eta})^{(-\varepsilon)} \}\rangle 
\langle
T_K\{c_{\sigma}(t^{\eta})^{(\varepsilon)}c_{\sigma}(t'^{-\eta})^{(-\varepsilon)}\}\rangle
~,\label{intermediate_tunneling_noise}
\end{eqnarray}
where the last factor in Eqs. (\ref{intermediate_tunneling_current}) and
(\ref{intermediate_tunneling_noise}) is the tip fermion Green's function.
Next the nanotube and tip fields are specified in
terms of the bosonized fields (nonchiral and chiral), and the two
Keldysh ordered exponential products are computed:
\begin{eqnarray}
\langle I_T(t)\rangle&=&
\frac{e\Gamma^2}{2(2\pi a)^2}\sum_{r\alpha\sigma\varepsilon\eta\eta_1}\eta\varepsilon
\int_{-\infty}^{+\infty}dt_1e^{-i\varepsilon\omega_0(t-t_1)}e^{2\pi g_{\sigma(\eta\eta_1)}(t-t_1)}\nonumber\\
&&\times
e^{\frac{\pi}{2}\sum_{j\delta}(G^{\phi\phi}_{j\delta(\eta\eta_1)}(0,0,t-t_1)
+rG^{\phi\theta}_{j\delta(\eta\eta_1)}(0,0,t-t_1)+rG^{\theta\phi}_{j\delta(\eta\eta_1)}(0,0,t-t_1)
+G^{\theta\theta}_{j\delta(\eta\eta_1)}(0,0,t-t_1))}~,\label{tunneling_current_theta_phi}\\
S_T(t,t')&=&
\frac{e^2\Gamma^2}{(2\pi a)^2}\sum_{r\alpha\sigma\varepsilon\eta}
e^{-i\varepsilon\omega_0(t-t')}
e^{2\pi g_{\sigma(\eta-\eta)}(t-t')}\nonumber\\
&&\times
e^{\frac{\pi}{2}\sum_{j\delta}(G^{\phi\phi}_{j\delta(\eta
-\eta)}(0,0,t-t')
+rG^{\phi\theta}_{j\delta(\eta
-\eta)}(0,0,t-t')+rG^{\theta\phi}_{j\delta(\eta
-\eta)}(0,0,t-t')
+G^{\theta\theta}_{j\delta(\eta-\eta)}(0,0,t-t'))}
~.\label{tunneling_noise_theta_phi}
\end{eqnarray}  
As expected, the stationary
current and the real time current correlator call for
the time differences $t-t_1$, $t-t'$ only. 
Integrating over time, the zero frequency noise  is introduced. 
Further using the symmetry properties
of the Green's functions
$g_{\sigma(\eta\eta)}(\tau)=g_{\sigma(\eta\eta)}(|\tau|)$ and
$G^{\phi\phi}_{j\delta(\eta\eta)}(0,0,\tau)=G^{\phi\phi}_{j\delta(\eta\eta)}(0,0,|\tau|)$ 
(similarly for $\theta\theta$, $\theta\phi$ and $\phi\phi$),
only $\eta=-\eta_1$ is retained for the current:
\begin{eqnarray}
\langle I_T\rangle&=&
-\frac{2ie\Gamma^2}{(2\pi a)^2}\sum_{r\sigma\eta}\eta \int_{-\infty}^{+\infty}d\tau 
\sin(\omega_0\tau)e^{2\pi g_{\sigma(\eta-\eta)}(\tau)}
e^{\frac{\pi}{2}\sum_{j\delta}(G^{\phi\phi}_{j\delta(\eta-\eta)}(0,0,\tau)
+rG^{\phi\theta}_{j\delta(\eta-\eta)}(0,0,\tau)+rG^{\theta\phi}_{j\delta(\eta-\eta)}(0,0,\tau)
+G^{\theta\theta}_{j\delta(\eta-\eta)}(0,0,\tau))}~,\nonumber\\
\label{general_tunneling_current}\\
S_T(\omega =0)&=&
-\frac{e^2\Gamma^2}{(\pi a)^2}\sum_{r\sigma\eta} 
\int_{-\infty}^{+\infty}d\tau \cos(\omega_0\tau)e^{2\pi
g_{\sigma(\eta-\eta)}(\tau)}
e^{\frac{\pi}{2}\sum_{j\delta}(G^{\phi\phi}_{j\delta(\eta-\eta)}(0,0,\tau)
+rG^{\phi\theta}_{j\delta(\eta-\eta)}(0,0,\tau)+rG^{\theta\phi}_{j\delta(\eta-\eta)}(0,0,\tau)
+G^{\theta\theta}_{j\delta(\eta-\eta)}(0,0,\tau))}~.\nonumber\\
\label{general_tunneling_noise} 
\end{eqnarray}  
The tunneling current and noise imply the
knowledge of the Green's functions at the tunneling location only. 

\subsection{Nanotube current and noise}

The operator averages along the nanotube 
require a perturbative calculation up to second order 
in the tunneling Hamiltonian for the tunneling current and
for the noise. Tunneling of an electron from the STM tip is
followed by propagation of the collective excitations of the
Luttinger liquid towards both ends of the nanotube.
\begin{eqnarray}
\langle I_\rho(x,t)\rangle&=&
-\frac{1}{4}\sum_{\eta\eta_1\eta_2}
\eta_1\eta_2 \langle T_K\{I_\rho(x,t^\eta)\int\int dt_1 dt_2
H_T(0,t_1^{\eta_1})H_T(0,t_2^{\eta_2})\}\rangle
~, \label{charge_current_average} \\
S_\rho(x,t;x',t') &=&-\frac{1}{4}\sum_{\eta\eta_1\eta_2}\eta_1\eta_2
\langle T_K\{I_\rho(x,t^\eta)I_\rho(x',t'^{-\eta}) \int\int
dt_1 dt_2 H_T(0,t_1^{\eta_1})H_T(0,t_2^{\eta_2})\}\rangle
~, \label{charge_noise_average}
\end{eqnarray}
where the contribution to the noise coming from
$\langle I_\rho(x,t)\rangle\langle I_\rho(x',t')\rangle$ has
been dropped because it contributes to order $\Gamma^4$. 
Expressing the Hamiltonian in terms of the fields, the
limit $\lim_{\gamma\to
0}(i\gamma)^{-1}\partial_x\exp[i\gamma
\phi_{c+}]=\partial_x \phi_{c+}$  is used in order to cast the
time ordered averages into  correlators of exponentials only: 
\begin{eqnarray} 
&&\langle I_\rho(x,t)\rangle=
-\frac{ev_F\Gamma^2}{4\pi
a}\sqrt{\frac{2}{\pi}}
\sum_{\eta\eta_1\eta_2\varepsilon_1r_1\alpha_1\sigma_1} \eta_1\eta_2 \int\int dt_1dt_2e^{-i\varepsilon_1\omega_0(t_1-t_2)}
\langle T_K\{c_{\sigma_1}^{(-\varepsilon_1)}(t_1^{\eta_1})
c_{\sigma_1}^{(\varepsilon_1)}(t_2^{\eta_2})\}\rangle\nonumber\\
&&\times\lim_{\gamma\rightarrow
0}\frac{1}{i\gamma}\partial_x\langle
T_K\{e^{i\gamma\phi_{c+}(x,t^\eta)}e^{-i\varepsilon_1\varphi_{r_1\alpha_1\sigma_1}(0,t_1^{\eta_1})}
e^{i\varepsilon_1\varphi_{r_1\alpha_1\sigma_1}(0,t_2^{\eta_2})}\}\rangle~, 
\label{nano_current_2}\\
&&S_\rho(x,t;x',t')=-\frac{e^2v^2_F\Gamma^2}{\pi^2 a}
\sum_{\eta\eta_1\eta_2\varepsilon_1r_1\alpha_1\sigma_1}\eta_1\eta_2
\int\int dt_1dt_2e^{-i\varepsilon_1\omega_0(t_1-t_2)}
\langle T_K\{c_{\sigma_1}^{(-\varepsilon_1)}(t_1^{\eta_1})
c_{\sigma_1}^{(\varepsilon_1)}(t_2^{\eta_2})\}\rangle\nonumber\\
&&\times\lim_{\gamma\rightarrow 0}
\frac{1}{\gamma^2}\partial_x\partial_{x'}
\langle
T_K\{e^{i\gamma\phi_{c+}(x,t^\eta)}
e^{-i\gamma\phi_{c+}(x',t'^{-\eta})} 
e^{-i\varepsilon_1\varphi_{r_1\alpha_1\sigma_1}(0,t_1^{\eta_1})}
e^{i\varepsilon_1\varphi_{r_1\alpha_1\sigma_1}(0,t_2^{\eta_2})}\}\rangle~,
\label{nano_noise_2}
\end{eqnarray}
where the contribution from the STM tip is the same as before.
The two time ordered products (one for the tip and one for the 
nanotube) are expressed in terms of 
Luttinger liquid  Green's functions. Taking
the spatial derivative, one obtains an expression with Green's
functions as prefactors -- implying propagation -- as well as
exponentiated Green's functions at the tunneling location.
Operating variable changes in the integrals and noticing that
only $\eta_1=-\eta_2$ contributes, the current  and noise become:
\begin{eqnarray} &&\langle I_\rho(x)\rangle=
-\frac{ev_F\Gamma^2}{2\pi^2 a^2}\sum_{\eta\eta_1r_1\sigma_1}
\int_{-\infty}^{+\infty}d\tau'\partial_x\left(G^{\phi\phi}_{c+(\eta\eta_1)}(x,0,\tau')
-G^{\phi\phi}_{c+(\eta-\eta_1)}(x,0,\tau')+r_1G^{\phi\theta}_{c+(\eta\eta_1)}(x,0,\tau')
-r_1G^{\phi\theta}_{c+(\eta-\eta_1)}(x,0,\tau')\right)\nonumber\\
&&\times \int_{-\infty}^{+\infty}d\tau sin(\omega_0\tau)e^{2\pi
g_{\sigma_1 (\eta_1-\eta_1)}(\tau)}
e^{\frac{\pi}{2}\sum_{j\delta}(G^{\phi\phi}_{{j\delta}(\eta_1-\eta_1)}(0,0,\tau)
+G^{\theta\theta}_{{j\delta}(\eta_1-\eta_1)}(0,0,\tau)
+r_1G^{\phi\theta}_{{j\delta}(\eta_1-\eta_1)}(0,0,\tau)
+r_1G^{\theta\phi}_{{j\delta}(\eta_1-\eta_1)}(0,0,\tau))}~,
\label{general_nanotube_current}\\
&&S_\rho(x,x',\omega=0)=-\frac{e^2v^2_F\Gamma^2}{(\pi a)^2}\sum_{\eta\eta_1r_1\sigma_1}\nonumber\\
&&\times \int_{-\infty}^{+\infty}d\tau cos(\omega_0\tau)
e^{2\pi g_{\sigma_1 (\eta_1-\eta_1)}(\tau)}
e^{\frac{\pi}{2}\sum_{j\delta}
(G^{\phi\phi}_{j\delta(\eta_1-\eta_1)}(0,0,\tau)+r_1G^{\phi\theta}_{j\delta(\eta_1-\eta_1)}(0,0,\tau)
+r_1G^{\theta\phi}_{j\delta(\eta_1-\eta_1)}(0,0,\tau)
+G^{\theta\theta}_{j\delta(\eta_1-\eta_1)}(0,0,\tau))}\nonumber\\
&&\times\int_{-\infty}^{+\infty}d\tau_1\partial_x\left(G^{\phi\phi}_{c+(\eta\eta_1)}(x,0,\tau_1)
-G^{\phi\phi}_{c+(\eta-\eta_1)}(x,0,\tau_1)+r_1G^{\phi\theta}_{c+(\eta\eta_1)}(x,0,\tau_1)
-r_1G^{\phi\theta}_{c+(\eta-\eta_1)}(x,0,\tau_1)\right)
\nonumber\\
&&\times\int_{-\infty}^{+\infty}d\tau_2\partial_{x'}\left
(G^{\phi\phi}_{c+(-\eta\eta_1)}(x',0,\tau_2)-G^{\phi\phi}_{c+(-\eta-\eta_1)}(x',0,\tau_2)
+r_1G^{\phi\theta}_{c+(-\eta\eta_1)}(x',0,\tau_2)
-r_1G^{\phi\theta}_{c+(-\eta-\eta_1)}(x',0,\tau_2)\right)
~.
\label{general_nanotube_noise}
\end{eqnarray}
Note the temporal decoupling (which occurs after operating
variable changes) in these expressions. The integral over
$\tau$ contains information on electron tunneling at $x=0$,
while the remaining integrals involve propagation, thus the
spatial dependence in the Green's functions arguments.  

\section{Current and noise for an infinite nanotube}  

In the previous section, general expressions were derived
for the current and noise, which are independent of the 
form of the Green's functions $G^{\phi\phi}_{j\delta}$,
$G^{\phi\theta}_{j\delta}$, $G^{\phi\theta}_{j\delta}$
and $G^{\theta\theta}_{j\delta}$.
The Green's functions
are described in Appendix A and are used 
to compute the tunneling noise and current as well as the
nanotube noise and current. 

\subsection{Tunneling current and noise}

After substitution of the Green's function of a nanotube, the
tunneling current and noise read:

\begin{eqnarray}
\langle I_T\rangle&=&
-\frac{2ie\Gamma^2}{(2\pi a)^2}\sum_{r\sigma \eta} \eta
\int_{-\infty}^{+\infty}\frac{\sin(\omega_0\tau)d\tau}
{(1-i\eta\frac{u_F^{\sigma}\tau}{a})(1-i\eta\frac{v_F\tau}{a})^\nu}~,
\label{current_green_explicited}\\
S_T(\omega=0)
&=&\frac{e^2\Gamma^2}{(\pi
a)^2}\sum_{r\sigma\eta}\int_{-\infty}^{+\infty}
\frac{\cos(\omega_0\tau)d\tau}{(1-i\eta\frac{u_F^\sigma}{a}\tau)
(1-i\eta\frac{v_F}{a}\tau)^\nu}
~,\label{noise_green_explicited}
\end{eqnarray}
with the exponent:
\begin{eqnarray}
\nu&=&\frac{1}{8}\sum_{j\delta}\left(K_{j\delta}+\frac{1}{K_{j\delta}}\right)~.
\label{exponents}
\end{eqnarray}
$\nu$ is the bulk tunneling exponent of the current--voltage
characteristics $\langle I_T (\omega_0)\rangle .^6$
The integrals are
computed in Appendix B, we obtain :
\begin{eqnarray}
\langle I_T\rangle&=&
\frac{2e\Gamma^2}{\pi a}\left(\sum_{\sigma}\frac{1}{u_F^{\sigma}}\right)
\left(\frac{a}{v_F}\right)^{\nu}\frac{\mathrm
{sgn}(\omega_0)|\omega_0|^{\nu}}{{\bf \Gamma}(\nu+1)}~,
\end{eqnarray} 
where we used the definition of the Gamma function ${\bf
\Gamma}$.
 Only electrons can tunnel
from  the tip to the nanotube, so one can check that the
classical Schottky  formula holds always:
\begin{equation}
S_T(\omega=0)=e|\langle I_T\rangle|~.
\end{equation}

\subsection{Nanotube current and noise}

Some of the time integrals in Eq.
(22) has already been encountered
when computing the tunneling current and noise. 
The current and noise thus become: 
\begin{eqnarray}
&& \langle I_\rho(x)\rangle=-\frac{eiv_F\Gamma^2}{\pi
a}\left(\sum_{\sigma}\frac{1}{u_F^{\sigma}}\right)
\left(\frac{a}{v_F}\right)^{\nu}\mathrm
{sgn}(\omega_0)\frac{|\omega_0|^{\nu}} {{\bf
\Gamma}(\nu+1)}\sum_{\eta\eta_1}\eta_1\int_{-\infty}^{+\infty}
d\tau'\partial_x\left(G^{\phi\phi}_{c+(\eta\eta_1)}(x,0,\tau')
-G^{\phi\phi}_{c+(\eta-\eta_1)}(x,0,\tau')\right)~,\nonumber\\
\label{current_1_integral}\\
&&
S_\rho(x,x',\omega=0)=
-\frac{2e^2v^2_F\Gamma^2}{\pi
a}\left(\sum_{\sigma}\frac{1}{u_F^{\sigma}}\right)
\left(\frac{a}{v_F}\right)^{\nu}\frac{|\omega_0|^{\nu}}{{\bf
\Gamma}(\nu+1)}\nonumber\\
&&\times\left[\sum_{\eta\eta_1}\int_{-\infty}^{+\infty}
d\tau_1\partial_x\left(G^{\phi\phi}_{c+(\eta\eta_1)}(x,0,\tau_1)
-G^{\phi\phi}_{c+(\eta-\eta_1)}(x,0,\tau_1)\right)
\int_{-\infty}^{+\infty}d\tau_2\partial_{x'}\left
(G^{\phi\phi}_{c+(-\eta\eta_1)}(x',0,\tau_2)-G^{\phi\phi}_{c+(-\eta-\eta_1)}
(x',0,\tau_2)\right)\right.\nonumber\\
&&\left.+\sum_{\eta\eta_1}\int_{-\infty}^{+\infty}d\tau_1
\partial_x\left(G^{\phi\theta}_{c+(\eta\eta_1)}(x,0,\tau_1)
-G^{\phi\theta}_{c+(\eta-\eta_1)}(x,0,\tau_1)\right)
\int_{-\infty}^{+\infty}d\tau_2\partial_{x'}\left
(G^{\phi\theta}_{c+(-\eta\eta_1)}(x',0,\tau_2)-G^{\phi\theta}_{c+(-\eta-\eta_1)}
(x',0,\tau_2)\right)\right]\nonumber\\
&&=-\frac{2e^2v^2_F\Gamma^2}{\pi
a}\left(\sum_{\sigma}\frac{1}{u_F^{\sigma}}\right)
\left(\frac{a}{v_F}\right)^{\nu}\frac{|\omega_0|^{\nu}}{{\bf
\Gamma}(\nu+1)}(I^{\phi\phi}(x,x')+I^{\phi\theta}(x,x'))~,
\nonumber\\
\end{eqnarray} 
where the last factors are computed in Appendix B.
The standard assumptions of the calculation of the tunneling
current and noise are recalled, as the same expressions appear in
both results. We obtain:  
\begin{eqnarray} \langle
I_\rho(x)\rangle&=&\frac{e\Gamma^2}{\pi
a}\left(\sum_{\sigma}\frac{1}{u_F^{\sigma}}\right)
\left(\frac{a}{v_F}\right)^{\nu}\frac{|\omega_0|^{\nu}\mathrm
{sgn}(\omega_0)}{{\bf \Gamma}(\nu+1)}\mathrm{sgn}(x)~,
\label{final_nanotube_current} \\
S_\rho(x,x',\omega=0)&=&\frac{(K_{c+})^2+\mathrm
{sgn}(x)\mathrm{sgn}(x')}{2} 
e|\langle I_\rho(x)\rangle|
~. \label{final_nanotube_noise}
\end{eqnarray} 

Current conservation
$|\langle I_\rho(x)\rangle|=
|\langle I_T\rangle| /2 $ is shown to hold. Results are valid for arbitrary 
voltages, with the expected power law behavior. 

\section{Discussion}

\subsection{Local current correlations} 

One accepted diagnosis to detect effective or anomalous charges
is to compare the noise with the associated current with the
Schottky formula in mind. A striking
result  is that despite the fact that electrons are tunneling 
from the STM tip to the bulk of the nanotube, the zero frequency
current fluctuations are proportional to the current 
for $x'=x>>a$:
\begin{eqnarray}
S_\rho(x,x,\omega=0)=
\frac{1+(K_{c+})^2}{2}e| \langle I_{\rho}(x)\rangle |~,
\end{eqnarray} 
with an anomalous effective charge for an infinite nanotube.

\subsection{Positive cross-correlations}

More can be learned from a measurement of the noise 
correlations. Noise correlations have been proposed to detect 
statistical correlations in quantum transport
\cite{martin_landauer_buttiker,safi_devillard_martin}. 
Indeed, our geometry can be considered as a Hanbury-Brown
and Twiss correlation device. Such experiments have now been
completed for photons and more recently for electrons in
quantum waveguides. Here the novelty is 
that electronic excitations do not represent the right 
eigenmodes of the nanotube. For $x'=-x>>a$ the noise
correlations read:
\begin{eqnarray}
S_\rho(x,-x,\omega=0)=-\frac{1-(K_{c+})^2}{2} 
e|\langle I_\rho(x)\rangle |~.
\label{positive_noise_correlations}
\end{eqnarray}
This is a priori negative. 
However, if the current direction is chosen to be positive from
the tip to the extremities of the nanotube, the sign of the 
cross--correlations is positive.
 Recall that the fermionic
version of the Hanbury-Brown and Twiss experiment yields 
negative noise correlations
\cite{martin_landauer_buttiker,henny}.  So far, positive noise
correlations have been attributed  in priority to bosonic systems
\cite{hbt}. Nevertheless, there are at least two other situations where
they are encountered. First, when the source of particle is a
superconductor, noise correlations  can also be positive
depending on the junction 
configuration\cite{torres_martin,gramespacher,belzig,schechter_taddei}. 
Second, they also occur in systems with floating voltage probes
\cite{texier}.  In the case of a superconductor, 
the emission of electron pairs
through separate quantum dots guarantee that 
the noise correlations are always positive: a (singlet) entangled 
electron pair is generated outside the superconductor 
\cite{recher_sukhorukov_loss,lesovik_martin_blatter}.

Note that the prefactors in Eq.
(\ref{positive_noise_correlations}) can readily be interpreted
using the language of Ref. \cite{ines_annales,pham,imura}. 
A tunneling event to the bulk of a
nanotube is accompanied by the propagation of two
counter-propagating charges $Q_\pm=(1\pm K_{c+})/2$. 
Recall that the subscript $c+$ identifies the charge
(as opposed to spin) excitation given by the total
(rather than relative) contribution of the two modes 
propagating in the nanotube.  Each
charge is as likely  to go right or left.  
According to Ref. \cite{pham} electron injection
in a Luttinger liquid is characterized by 
chiral charges $Q_\pm$ and chiral spin charges 
$S_\pm$ which describe the elementary excitations of the
nanotube. 
\begin{equation} 
\left( \begin{array}{ll} Q_+\\
Q_-\\
S_+\\
S_-
\end{array}
\right)
=\sum_\sigma  
\left[
n_\sigma
\left(
\begin{array}{ll}
1\\
1\\
\sigma/2 \\
\sigma/2
\end{array}
\right)
+J_\sigma
\left(
\begin{array}{ll}
(1+K_{c+})/2\\
(1-K_{c+})/2\\
\sigma(1+K_{s+})/4 \\
\sigma(1-K_{s+})/4
\end{array}
\right)
\right]~,
\end{equation}
with integers $n_\sigma,J_\sigma=0,1,2,...$
($\sigma=\uparrow,\downarrow$). In particular, 
the addition of an electron  with spin $\sigma$ corresponds to
the choice $n_\sigma=0$ and $J_\sigma=1$. 

The current noise and noise correlations can
be interpreted as  an average over the two types of excitations:
\begin{eqnarray}
S_\rho (x,x) &\sim& {( Q_+^2+Q_-^2 )\over 2} ={1+(K_{c+})^2 \over
4}~,\label{noise_quasiparticle}\\ 
S_\rho (x,-x) &\sim& -Q_+Q_- =-{1-(K_{c+})^2 \over
4}~.\label{correlations_quasiparticle}
\end{eqnarray}

\begin{figure}[h]  
\epsfxsize 8 cm  
\centerline{\epsffile{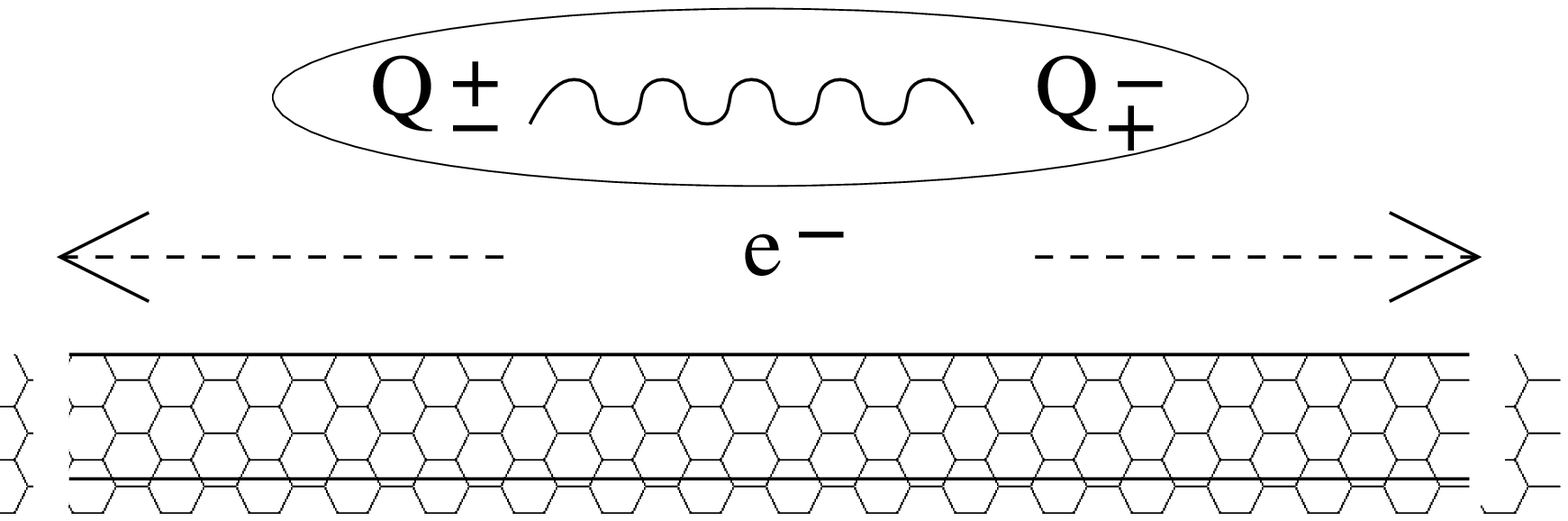}}  
\medskip  
\caption{\label{fig2} Schematic description of entangled quasiparticles $Q_\pm$ being emitted right
and left of the tunnel junction.}    
\end{figure} 

A drawing where the two types of charges ``flow away''
from the tip while propagating along the nanotube 
is depicted in the lower part of Fig. \ref{fig2}. Both charges $Q_\pm$ are equally
likely to go right or left, and they are emitted as a pair with opposite labels. 
The noise correlations of Eq.(\ref{correlations_quasiparticle})
are rendered positive if one adopts the standard convention for measuring the current in 
multi-terminal conductors \cite{blanter_buttiker}.
Here these ``positive'' noise correlations resulting from charges 
moving toward both extremities of the nanotube
have the added particularity  that they occur to second order in 
a perturbative tunneling calculation. 
In superconducting-normal systems, the two electrons which emanate from the
same Cooper pair
and which propagate in the two Luttinger liquids provide a 
manifestation of the non local character of quantum mechanics. 
In the present case, only one electron  is injected, but it is
split into left and right excitations, unless one imposes one
dimensional Fermi liquid leads. Here, we are dealing with 
entanglement between collective excitations 
of the Luttinger liquid. 
Written in terms of the chiral quasiparticle fields, the addition
of an electron with given spin $\sigma$ on a nanotube in the
ground state $|O_{LL}\rangle$ gives: 
\begin{equation} 
\sum_{r\alpha}
\Psi^{\dag}_{r\alpha\sigma}(x=0)|O_{LL}\rangle =
{1\over \sqrt{2\pi a}}
\sum_{r\alpha}
\exp
\left[-
i\sum_{j\delta}
\sqrt{\frac{\pi}{2 K_{j\delta}}}
h_{\alpha\sigma j\delta}
\left(
\frac{1+rK_{j\delta}}{2}\tilde{\varphi}_{j\delta}^+(x)
+\frac{1-rK_{j\delta}}{2}\tilde{\varphi}_{j\delta}^-(x)
\right)
\right]
|O_{LL}\rangle~,
\label{fermion_nanotube_entanglement}
\end{equation}
with $\tilde{\varphi}_{j\delta}^r$ the chiral bosonic fields of
the (nonchiral) Luttinger liquid.  
This wave function is
characterized by right and left movers $r=\pm$  whose fields
appear explicitly in the phase operator of this many-particle 
wave function. These fields are independent of each other, therefore the
exponential can be written as a product of fields:
\begin{equation}
\sum_{r\alpha}\Psi^{\dag}_{r\alpha\sigma}(x=0)|O_{LL}\rangle =
{1\over \sqrt{2\pi a}}\sum_{\alpha}
\prod_{j\delta}
\left[
(\tilde{\psi}^{\dag}_{j\delta +})^{Q_{j\delta +}}(\tilde{\psi}^{\dag}_{j\delta -})^{Q_{j\delta -}}
+(\tilde{\psi}^{\dag}_{j\delta +})^{Q_{j\delta -}}(\tilde{\psi}^{\dag}_{j\delta -})^{Q_{j\delta +}}
\right]
|O_{LL}\rangle~,
\label{entangled_wave_function}
\end{equation}
where for each sector (charge/spin, total/relative mode) the charges
$Q_{j \delta \pm}=(1\pm K_{j \delta})/2$ have been introduced, and 
chiral fractional operators are defined as: 
\begin{equation}
\tilde{\psi}_{j\delta \pm}(x)=
\exp
\left[i 
\sqrt{\frac{\pi }{2 K_{j\delta}}}
h_{\alpha\sigma j\delta}\tilde{\varphi}_{j\delta}^\pm(x)
\right]~.
\label{quasiparticle_chiral}
\end{equation}
The wave function described by Eq.
(\ref{entangled_wave_function}) has all the characteristics 
of an entangled state. Because the two types of excitations 
travel towards opposite ends of the nanotube, the time evolution
of this ``injected electron '' state is simply obtained with the
substitution
$\tilde{\varphi}_{j\delta}^r(x)\to\tilde{\varphi}_{j\delta}^r(x-rv_{j\delta}t)$. 
Consequently, quantum mechanical non-locality is quite explicit
here. The detection of a charge $Q_\pm$ in one arm is necessarily
accompanied by the simultaneous detection of a charge $Q_\mp$ in
the other extremity of the nanotube.  

This
entanglement is the direct consequence of the correlated
state of the Luttinger liquid. When additional electrons are
injected, these break up into the specific modes which can
propagate in either direction in the nanotube. It therefore
differs significantly from its  analogs which use
superconductors as electron injectors,  where two electrons from
the same Cooper pair are dissociated
\cite{recher_sukhorukov_loss,lesovik_martin_blatter,%
recher_loss,bena,bouchiat}. 

When considering only one sector, such as $j\delta=c+$, it is interesting to note that
the wave function has the same structure of say, a triplet spin state
(a symmetric combination of ``up'' and ``down'' states, or
``plus'' and ``minus'' charges) for electrons, with the electrons
being replaced by chiral quasiparticle operators. 
Indeed, one has to recognize that each chiral field
$\tilde{\varphi}_{j\delta}^r$ can be written as a superposition
of boson operators:
\begin{equation}
\tilde{\varphi}_{j\delta}^{r}(x)=
{1\over 4\sqrt{K_{j\delta}}}
\sum_{r'\alpha \sigma}h_{\alpha\sigma j\delta}(r+r'K_{j\delta})
\sum_{(r'k)>0}\sqrt{\frac{1}{|k|L}}
\left( d^{\dagger}_{\alpha\sigma}(k) e^{-ikx}+d_{\alpha\sigma}(k) e^{ikx}
\right) e^{-a|k|/2}~,
\label{bosons_chiral}
\end{equation}
where $d^{\dagger}_{\alpha\sigma}(k)$ creates a boson with nanotube mode
$\alpha$, spin $\sigma$ and momentum $k$, and characterizes the collective 
modes of the one dimensional liquid. 
According to the state written in Eq. (\ref{quasiparticle_chiral}),
this linear superposition of boson operators appears in an exponential. This
expresses that non-local ``many--boson'' correlations are
created when an electron is injected in a nanotube, and these 
many-body states are entangled in the present geometry.

\subsection{Spin current}

Effects similar to the detection of effective charges show up in the spin sector when time reversal symmetry ($K_{s+}\ne 1$) does not hold. The spin current and
spin noise are obtained in a similar manner:
\begin{eqnarray}
\langle I_{\sigma_z}(x)\rangle&=&
\frac{e\Gamma^2}{\pi
a}\bigg(\sum_{\sigma}\frac{\sigma}{u_F^{\sigma}}\bigg)
\frac{{\mathrm sgn}(\omega_0)|\omega_0|^{\nu}}{{\bf
\Gamma}(\nu+1)}\bigg(\frac{a}{v_F}\bigg)^{\nu}
{\mathrm sgn}(x)~.
\label{result_spin_current}
\end{eqnarray}
So that at large distances:
\begin{eqnarray}
S_{\sigma_z}(x,-x,\omega=0)&=&-\frac{1-(K_{s+})^2}{2} e
|\langle I_{\sigma_z}(x)\rangle|~,
\label{spin_noise_infinite}\\
S_{\sigma_z}(x,x,\omega=0)&=&
\frac{1+(K_{s+})^2}{2}e|
\langle I_{\sigma_z}(x)\rangle |~. 
\label{spin_correlations_infinite}
\end{eqnarray}
In practice, when time
reversal symmetry holds ($K_{s+}=1$), spin noise correlations vanish to order
$\Gamma^2$ independently from the presence or 
the nature of the leads. In the case where the tip is
non magnetized, the spin current and  spin noise correlations
also vanish.

\subsection{1D Fermi liquid leads}

In the presence of one-dimensional Fermi liquid leads, where the leads are considered to be Luttinger liquids
whose interaction parameters are set to $K_{j\delta}^L\equiv 1$, quasiparticles suffer Andreev
type reflections \cite{ines_annales} at both exterminites of the nanotubes. 
Multiple reflections of quasiparticles in the Fabry-Perot geometry -- Fermi liquid/Nanotube/Fermi
liquid -- are expected to lead to a concellation of the interaction effects in the nanotube,
as in the two terminal calculations of conductance and noise \cite{safi_maslov}. 
Although the detailed calculation is not presented here, dimensional analysis of the time integrals suggest that,
the nanotube current and noise read:
\begin{eqnarray}
\langle I_\rho(x) \rangle &=& \frac{e\Gamma^2 \omega_0}{\pi
v_F} \left(\sum_{\sigma}\frac{1}{u_F^{\sigma}}\right){\mathrm
sgn}(x)~, \label{fermi_liquid_current}
\\
S_\rho(x,x',\omega=0)&=& \frac{1+{\mathrm sgn}(x){\mathrm
sgn}(x')}{2} e|\langle
I_\rho(x)\rangle|~.
\label{fermi_liquid_noise}
\end{eqnarray}
For $x=x'$, this whould give the classical Schottky
formula, in the very same spirit as in Ref. \cite{safi_maslov}.
For $x$ and $x'$ on opposite ends of the nanotube, 
this noise correlator should vanish, to this order:
the scattering theory result has a lowest non vanishing
contribution of order $\Gamma^4$.

This low voltage result is
modified by a higher power law behavior at higher voltage, with
a threshold voltage specified by the size of the system
$\hbar v_F/L$ as in Ref. \cite{safi_maslov}.

\section{Conclusion}

In summary, a diagnosis for detecting the chiral excitations of 
a Luttinger liquid nanotube has been presented, which is based
on the knowledge of low frequency current fluctuation spectrum 
in the nanotube. Typical transport calculations either address
the propagation in a nanotube, or compute tunneling I(V)
characteristics. Here, both are addressed because they
constitute the key for obtaining the quasiparticle charges. 
Both the noise (autocorrelation) and the noise
correlations (cross-correlations) are needed to identify the charges $Q_{\pm}$. 
Independently, note that this measurement could also be 
confronted to other diagnoses of the nanotube interaction
parameter using tunneling current voltage characteristics.

This result relies on the
assumption that one dimensional Fermi liquid leads are avoided.
Such leads have been treated in different approaches
\cite{safi_euro,egger_grabert}, and are also labelled radiative
contacts. Radiative contacts imply equilibration with the
electrons.
In Ref. \cite{imura}, both radiative contacts and equilibration with
dressed eigenmodes were studied, with the obvious result that 
Luttinger liquid renormalization shows up in the conductance
in the latter case. 
In special
circumstances such as the case of Ref. \cite{roche}
the nanotube is embedded in the metallic contacts, 
and it is suspended by its ends. Here, the absence of a 
screening gate is explicit. Electron transport 
between these two entities likely occurs in multiple electron
scattering processes as studied in Ref. \cite{chamon_fradkin}.
In these, or other contacts fabricated by growing techniques
\cite{bouchiat}, quantities such as current and noise may not
be affected by the presence of the contacts. 

Standard fermion results should be recovered when the system is
connected on one dimensional Fermi liquid leads. The
auto-correlation noise in one end of the nanotube should be related 
to the charge current with the standard Schottky formula.
The noise correlation signal should also vanish as expected and
the next order correction $O(\Gamma^4)$ then needs to be
computed. 

A crucial test of the contacts is in order. It should be possible
in practical situations to analyze the type of contacts which one
has between the nanotube and its connections.  If the ratio
of the cross-correlations to the current
$S_\rho (x,-x,\omega_0)/\langle I_\rho(x)\rangle$ does not depend
on the tunneling distance ($\log \Gamma$), both contributions are
of order $\Gamma^2$ and this constitutes an indication that the 
contacts do not  affect this quasiparticle entanglement. If we
are dealing with a Fermi liquid behavior,  the 
noise correlation--current ratio should behave like $\Gamma^2$, rather than
a constant.

Finally, we have remarked that the many-body wave function
which describes  a Luttinger liquid with an added electron  has
necessarily  EPR \cite{epr} entangled degrees of freedom. 
Both electrons chiralities contribute to the emission
of quasiparticle pairs moving in opposite direction.
This entanglement involves many particle states, unlike its
electron counterpart. A suggestion
for detection of such Luttinger liquid entanglement
without perturbing the system with leads is nevertheless needed. 
The issue -- how to detect this many-body 
entanglement -- should be addressed while taking into
account different models for the leads, possibly involving
multiple reflections within one contact \cite{chamon_fradkin}.
Multiple reflections of the quasiparticles
from one contact to the other kill this entanglement 
in one dimensional Fermi liquid leads, which is implicit 
in the vanishing of the noise correlations to this order.
At any rate, this is the first time that collective excitations 
entanglement is discussed in a condensed matter setting.

\acknowledgements
Discussions with A. Lebedev, I. Safi on the Keldysh Green's function and with
M. B\"uttiker are gratefully
acknowledged. 

\appendix
\section{Green's functions in the presence of contacts}
\label{green_appendix}

In this appendix, the Green's functions are computed, assuming 
a Luttinger liquid with an homogeneous interaction
parameter $K_{j\delta}$ and velocity $v_{j\delta}$. The product $v_{j\delta}K_{j\delta}$
corresponds to the Fermi velocity $v_F$.

The finite temperature action associated with this problem has
the general form: 
\begin{eqnarray}
S=\frac{1}{2}\sum_{j\delta}\int_{0}^{\beta}d\tau\int_{-\infty}^{\infty}dx
\bigg(v_{j\delta}K_{j\delta}(\partial_x\phi_{j\delta}(x,\tau))^2
+\frac{v_{j\delta}}{K_{j\delta}}(\partial_x\theta_{j\delta}(x,\tau))^2
+2i(\partial_x\phi_{j\delta}(x,\tau))(\partial_\tau\theta_{j\delta}(x,\tau))\bigg)~.
\label{action}
\end{eqnarray}
Which implies that the Fourier transform
of the time ordered Green's functions $G^{\theta\theta}_{j\delta(T)}$ 
and $G^{\phi\phi}_{j\delta(T)}$, define as:
\begin{eqnarray}
G^{\theta\theta}_{j\delta(T)}(x,x',t)&=&
\langle T\theta_{j\delta}(x,t)\theta_{j\delta}(x',0)\rangle
-\langle T\theta_{j\delta}^2(x,t)\rangle~,\\
G^{\phi\phi}_{j\delta(T)}(x,x',t)&=&
\langle T\phi_{j\delta}(x,t)\phi_{j\delta}(x',0)\rangle
-\langle T\phi_{j\delta}^2(x,t)\rangle~,
\end{eqnarray}

where T is the time ordered operator, satisfies the differential equations$^{12}$: 
\begin{eqnarray}
&&\bigg(\frac{\omega^2}{v_{j\delta}K_{j\delta}}
-\partial_x\frac{v_{j\delta}}{K_{j\delta}}\partial_x\bigg)
G^{\theta\theta}_{j\delta(T)}(x,x',\omega)=4\pi\delta(x-x')~,\label{green_theta}\\
&&\bigg(-\frac{K_{j\delta}\omega^2}{v_{j\delta}}
+\partial_xv_{j\delta}K_{j\delta}\partial_x\bigg)
G^{\phi\phi}_{j\delta(T)}(x,x',\omega)=4\pi\delta(x-x')~.\label{green_phi}
\end{eqnarray}
The Green's function $G^{\theta\theta}_{j\delta(T)}$ is continuous
everywhere, and
$v_{j\delta}[\partial_x G^{\theta\theta}_{j\delta(T)}]/K_{j\delta}$ has a
discontinuity at $x=x'$. The similarity between Eq. 
(\ref{green_theta}) and  (\ref{green_phi}) results from the 
duality properties of the underlying fields. All
information on $G^{\phi\phi}_{j\delta(T)}$ is obtained by dividing 
$G^{\theta\theta}_{j\delta(T)}$ by $K_{j\delta}^2$.

According to Eqs. (14) and
(15), there are additional Green's
functions in our problem which involve the fields $\theta$ and
$\phi$. For instance:
\begin{eqnarray}
G^{\phi\theta}_{j\delta(T)}(x,x',t)=
\langle T\phi_{j\delta}(x,t)\theta_{j\delta}(x',0)\rangle
-\langle T\phi_{j\delta}(x,t)\theta_{j\delta}(x,t)\rangle~.
\end{eqnarray}
Using the action (\ref{action}) one can show that ($t$ is a
real time variable): 
\begin{eqnarray}
\langle\partial_x\phi_{j\delta}(x,t)\theta_{j\delta}(x',0)\rangle
=\frac{1}{v_{j\delta}K_{j\delta}}\langle\partial_t\theta_{j\delta}(x,t)
\theta_{j\delta}(x',0)\rangle~,
\end{eqnarray}
and similarly for $G^{\theta\phi}_{j\delta(T)}$.

From these real time Green's functions, we further specify
the Keldysh matrix elements which two times $t,0$ are assigned to
the upper/lower branch ($++,+-,-+,--$). Given an arbitrary real
time Green's function $G(x,x',t)=\langle A(x,t)B(x',0)\rangle-\langle A(x,t)B(x,t)\rangle$
a general procedure$^{39}$ for obtaining these
elements is as follows:
\begin{eqnarray}
G^{\theta\theta}_{j\delta(K)}(x,x',t)=
\left(\begin{array}{cc}
G^{\theta\theta}_{j\delta}(x,x',|t|)
&G^{\theta\theta}_{j\delta}(x',x,-t)\\
G^{\theta\theta}_{j\delta}(x,x',t)&
G^{\theta\theta}_{j\delta}(x',x,-|t|)
\end{array}\right)~,
\end{eqnarray}
where:
\begin{eqnarray}
G^{\theta\theta}_{j\delta}(x,x',t)=~-\frac{K_{j\delta}}{8\pi}
\sum_r\mathrm{ln}\left(1+i\frac{v_Ft}{a}+ir\frac{K_{j\delta}(x-x')}{a}\right)~.
\end{eqnarray}
The same applies to $G^{\phi\phi}_{j\delta(K)}$ for which we have:
\begin{eqnarray}
G^{\phi\phi}_{j\delta}(x,x',t)=~-\frac{1}{8\pi K_{j\delta}}
\sum_r\mathrm{ln}\left(1+i\frac{v_Ft}{a}+ir\frac{K_{j\delta}(x-x')}{a}\right)~.
\end{eqnarray}

The mixed correlators read: 
\begin{eqnarray}
G^{\phi\theta}_{j\delta(K)}(x,x',t)=
\left(\begin{array}{cc}
\begin{array}{ll}
t>0 : G^{\phi\theta}_{j\delta}(x,x',t)\\
t<0 : G^{\theta\phi}_{j\delta}(x',x,-t)
\end{array}
&G^{\theta\phi}_{j\delta}(x',x,-t)\\
G^{\phi\theta}_{j\delta}(x,x',t)&
\begin{array}{ll}
t>0 : G^{\theta\phi}_{j\delta}(x',x,-t)\\
t<0 : G^{\phi\theta}_{j\delta}(x,x',t)
\end{array}
\end{array}\right)
~.
\end{eqnarray}
where:
\begin{eqnarray}
G^{\theta\phi}_{j\delta}(x,x',t)=~-\frac{1}{8\pi}
\sum_r r\mathrm{ln}\left(1+i\frac{v_Ft}{a}+ir\frac{K_{j\delta}(x-x')}{a}\right)~.
\end{eqnarray}
The same applies to $G^{\theta\phi}_{j\delta(K)}$ for which we have:
\begin{eqnarray}
G^{\phi\theta}_{j\delta}(x,x',t)=~-\frac{1}{8\pi}
\sum_r r\mathrm{ln}\left(1+i\frac{v_Ft}{a}+ir\frac{K_{j\delta}(x-x')}{a}\right)~.
\end{eqnarray}
\section{Integrals}
\label{appendix_integrals}

We now compute the integrals involved in the tunneling current
and noise. The general integrals which will be
required  to compute the current and noise read: 
\begin{eqnarray}
\int_{-\infty}^{+\infty}\frac{\sin(\omega_0\tau)d\tau}
{\left(\frac{a}{u^{\sigma}_F}-i\eta\tau\right)\left(\frac{a}{v_F}-i\eta\tau\right)^{\nu}}
\approx i\pi\eta\mathrm{sgn}(\omega_0)\frac{|\omega_0|^{\nu}}{{\bf
\Gamma}(\nu+1)} ~,\\
\int_{-\infty}^{+\infty}\frac{\cos(\omega_0\tau)d\tau}
{\left(\frac{a}{u^{\sigma}_F}-i\eta\tau\right)\left(\frac{a}{v_F}-i\eta\tau\right)^{\nu}}
\approx \pi\mathrm{sgn}(\omega_0)\frac{|\omega_0|^{\nu}}{{\bf
\Gamma}(\nu+1)} ~.
\end{eqnarray}

We now write the integral which 
appears in the nanotube current, which refer to propagation
along the nanotube: 
\begin{eqnarray}
I_2&=&\int_{-\infty}^{+\infty}d\tau'\partial_x\left(G^{\phi\phi}_{c+(++)}(x,0,\tau')
-G^{\phi\phi}_{c+(--)}(x,0,\tau')+G^{\phi\phi}_{c+(-+)}(x,0,\tau')
-G^{\phi\phi}_{c+(+-)}(x,0,\tau')\right)~.
\end{eqnarray}
Using the expressions for the Green's functions (Appendix \ref{green_appendix}): 
\begin{eqnarray}
I_2&=&\frac{i}{\pi v_F} \arctan\left(\frac{K_{j\delta}x}{a}\right)
\approx i  \frac{\mathrm{sgn}(x)}{2v_F}~,
\end{eqnarray} where the approximate sign holds at
large distances.

The integrals which are involved for the computation of the
noise read:
\begin{eqnarray}
I^{\phi\phi}(x,x')&=&4I_3(x)I_3(x')~,\\
I^{\phi\theta}(x,x')&=&4I_4(x)I_4(x')~,
\end{eqnarray}
with
\begin{eqnarray}
I_3(x)&=&\int_{-\infty}^{+\infty}d\tau\partial_{x}\left 
(G^{\phi\phi}_{c+(++)}(x,0,\tau)-G^{\phi\phi}_{c+(+-)}(x,0,t)\right)\approx i\;\mathrm{sgn}(x)/4v_F~,\\ 
I_4(x)&=&\int_{-\infty}^{+\infty}d\tau\partial_{x}\left
(G^{\phi\theta}_{c+(++)}(x,0,\tau)-G^{\phi\theta}_{c+(+-)}(x,0,t)\right)\approx -iK_{c+}/4v_F~. 
\end{eqnarray}


\end{document}